\documentclass[10pt,conference]{IEEEtran}
\IEEEoverridecommandlockouts

\usepackage{multirow}
\usepackage{graphicx}
\usepackage{booktabs}
\usepackage{subcaption}
\usepackage{float}
\usepackage{enumitem}
\usepackage{balance}

\usepackage{subcaption}
\usepackage{adjustbox}
\usepackage{caption}
\usepackage{url}
\usepackage{cite}
\usepackage{amsmath,amssymb,amsfonts}
\usepackage{algorithmic}
\usepackage{graphicx}
\usepackage{textcomp}
\usepackage{xcolor}
\usepackage{url}
\usepackage{balance}
\usepackage{cite}
\usepackage{amsmath,amssymb,amsfonts}
\usepackage{algorithmic}
\usepackage{graphicx}
\usepackage{textcomp}
\usepackage{xcolor}
\usepackage{float}
\usepackage{url}
\usepackage[colorlinks=true, urlcolor=blue, citecolor=black, linkcolor=black]{hyperref}

\usepackage[most]{tcolorbox}
\usepackage{csquotes}
\newtcolorbox{myquote}[1][]{%
    colback=black!5,
    colframe=black!5,
    notitle,
    sharp corners,
    borderline west={2pt}{0pt}{black!80!black},
    enhanced,
    breakable,
    top=0.5pt,
    bottom=0.5pt
}

\usepackage{array, tabularx}

\def\BibTeX{{\rm B\kern-.05em{\sc i\kern-.025em b}\kern-.08em
    T\kern-.1667em\lower.7ex\hbox{E}\kern-.125emX}}
\begin{document}

\title{Analyzing the Evolution and Maintenance of Quantum Software Repositories}

\newcommand{\stitle}[1]
{\noindent\textup{\textbf{#1}}}
\newcommand\myNum[1]{\emph{{(#1)}}}

\newcommand\todou[1]{\textcolor{red}{\textit{Umar: #1}}}
\newcommand\todok[1]{\textcolor{blue}{\textit{Krishna: #1}}}
\newcommand\todov[1]{\textcolor{blue}{\textit{Vinaik: #1}}}

\author{\IEEEauthorblockN{Krishna Upadhyay\IEEEauthorrefmark{1}, Vinaik Chhetri\IEEEauthorrefmark{1}, A.B. Siddique\IEEEauthorrefmark{2}, Umar Farooq\IEEEauthorrefmark{1}}
\IEEEauthorblockA{\IEEEauthorrefmark{1}
    Louisiana State University, 
    \IEEEauthorrefmark{2} 
    University of Kentucky \\
 Email: 
kupadh4@lsu.edu, vchhet2@lsu.edu, siddique@cs.uky.edu, ufarooq@lsu.edu
}}

\maketitle

\thispagestyle{plain}
\pagestyle{plain}
\begin{abstract}
Quantum computing is rapidly advancing, but quantum software development faces significant challenges, including a steep learning curve, high hardware error rates, and a lack of mature engineering practices. This study conducts a large-scale mining analysis of over 21,000 GitHub repositories, containing 1.2 million commits from more than 10,000 developers, to examine the evolution and maintenance of quantum software.  
We analyze repository growth, programming language and framework adoption, and contributor trends, revealing a 200\% increase in repositories and a 150\% rise in contributors since 2017. Additionally, we investigate software development and maintenance practices, showing that perfective commits dominate (51.76\%), while the low occurrence of corrective commits (18.54\%) indicates potential gaps in bug resolution. 
Furthermore, 34\% of reported issues are quantum-specific, highlighting the need for specialized debugging tools beyond conventional software engineering approaches.  
This study provides empirical insights into the software engineering challenges of quantum computing, offering recommendations to improve development workflows, tooling, and documentation. 
We are also open-sourcing our dataset to support further analysis by the community and to guide future research and tool development for quantum computing.
The dataset is available at: \url{https://github.com/kriss-u/QRepoAnalysis-Paper}.

\end{abstract}


\section{Introduction}
\label{sec:intro}
Quantum computing has the potential to transform industries by solving problems that are currently beyond the capabilities of classical computers. 
Quantum computers can perform calculations exponentially faster in certain applications using quantum mechanics and facilitate breakthroughs in many fields such as cryptography, material science, and optimization.
Shor's algorithm threatens modern encryption by efficiently factoring large integers~\cite{shor1999polynomial}, while Grover's algorithm accelerates unstructured database searches~\cite{grover1996fast}.
The economic impact is expected to be significant, with projections suggesting that the global quantum computing market could surpass \$125 billion by 2030~\cite{size2023share}. 
This potential has driven major investments, such as the \$1.2 billion U.S. National Quantum Initiative~\cite{us-quantum-initiative}, along with contributions from industry including IBM, Google, and Microsoft~\cite{companies-in-quantum}.
However, quantum software development faces unique challenges compared to classical computing, such as the probabilistic nature of quantum operations, high error rates in hardware, and a steep learning curve for developers~\cite{haner2018software}.

Despite rapid advancements in quantum hardware, quantum software development remains challenging and underdeveloped. Frameworks like Qiskit~\cite{qiskitprogramming2021} and Cirq~\cite{cirq} are still maturing and lack the robustness of classical software ecosystems~\cite{cross2022openqasm, qiskitprogramming2021}. 
Developers face a steep learning curve due to the complexity of quantum models, the probabilistic nature of computations, and the need for a strong grasp of quantum mechanics~\cite{corcoles2019challenges, larose2019overview}. 
The field also suffers from a lack of standardized tools, debugging environments, and educational resources, which collectively hinder progress~\cite{haner2018software}.
These challenges highlight the need for analyzing real-world repositories to better understand current quantum programming practices and identify key areas for improvement.

\stitle{State-of-the-Art.}
Prior research on quantum software has largely concentrated on small-scale or qualitative analyses of developer challenges. 
Paltenghi and Pradel \cite{paltenghi-quantum-study} examined 18 quantum projects to identify recurring bug patterns, while Zhao et al. \cite{zhao2023identifying} analyzed 36 bugs in Qiskit, highlighting key debugging challenges. 
In a separate study, Zhao and others \cite{zhao_qml} explored 391 quantum machine learning issues across 22 projects.
More recently, Chen et al. \cite{smellyeight} investigated quantum-specific code smells in 15 programs. 
Other studies have analyzed discussions from Stack Exchange, GitHub issues, and Xanadu Discussion Forums~\cite{quantum-issues, zappin2024quantum}, offering qualitative insights into quantum software engineering. 
However, these works are either constrained by their limited dataset size or focus on specific frameworks, leaving a gap in understanding the broader landscape of quantum software development.

\stitle{Overview of This Work.}
Unlike prior work, this work conducts a systematic mining analysis of over 21 thousand GitHub repositories, which contain more than 1.2 million commits contributed by more than 10,000 unique developers. 
By leveraging this large dataset, we extract broad trends, analyze software activity patterns, and identify key challenges in quantum computing development. 
This data-driven approach provides a comprehensive perspective on the state of quantum software and offers previously unexplored quantitative insights.
Our analysis reveals a 200\% increase in the number of repositories and a 150\% rise in contributors since 2017, indicating the rapid expansion of quantum computing thanks to the availability of frameworks like Qiskit~\cite{qiskitprogramming2021}, Cirq~\cite{cirq}, and PennyLane~\cite{pennylane}.
We further examine trends in programming languages and frameworks and find out that Python and Qiskit dominate but that there is also a shift toward specialized tools.
Additionally, we discover that quantum computing projects are focused on simulation, machine learning, and algorithm development, reinforcing the field's evolving priorities and the need for robust infrastructure and quantum-specific tools.

While analyzing the state of maintenance, we examine commit patterns and note an increase in both commit volume and size over time, which reflects growing project complexity and heightened developer engagement.
A deeper analysis of reported issues reveals that over 34\% of the issues are specific to quantum computing. 
Furthermore, our analysis of commits identifies that perfective commits, aimed at refining functionality, are the most common, while fewer corrective commits suggest areas for improvement in bug resolution.
Last but not least, we categorize repositories by activity levels and demonstrate that high-activity repositories typically feature frequent updates, larger codebases, and greater community involvement, in contrast to low-activity projects, which exhibit sporadic updates.

Our findings highlight challenges and opportunities in quantum software engineering, emphasizing the need for improved development workflows, enhanced debugging tools, and more structured maintenance practices. 
The dominance of perfective commits over corrective commits suggests a gap in bug resolution, indicating that better issue-tracking mechanisms and automated testing frameworks could strengthen software reliability. 
The high proportion of quantum-specific issues indicates the necessity for specialized tooling, such as domain-aware debuggers and enhanced simulators, to support developers in tackling the unique complexities of quantum programming. 
Moreover, the disparity between high- and low-activity repositories suggests that community-driven initiatives, improved documentation, and more accessible learning resources could help sustain engagement in quantum software projects.

In summary, this work establishes a solid foundation to drive targeted advancements in quantum software and to promote sustained growth and technical progress.
Through an in-depth analysis of development patterns, community structure, and maintenance practices, we provide actionable insights for enhancing quantum programming tools, documentation, and community resources.
We are open-sourcing our dataset to facilitate additional analysis by the community to guide future research and inspire the development of specialized tools for the unique demands of quantum computing.

\section{Background and Related Work}
\label{sec:background}
\subsection{Quantum Software Development}
Quantum software encompasses quantum platforms that provide the tools and environments developers need to write and execute quantum programs. 
Developers use these platforms to create algorithms that run on quantum hardware or simulators, bridging theoretical quantum concepts with practical applications.

Quantum platforms generally have three main components: a programming language, a compiler, and an execution environment. 
Quantum programming languages include both API-based libraries and stand-alone languages. 
For instance, Qiskit provides quantum programming abstractions as a Python library, making it accessible within a well-known host language~\cite{javadi2024quantum}. 
In contrast, other languages, such as Q\#~\cite{q-sharp}, Silq~\cite{silq}, and Quipper~\cite{green2013quipper}, are stand-alone languages specifically designed for quantum programming, offering dedicated syntax and semantics to better align with quantum computing paradigms. 
These languages help developers create complex quantum operations through high-level abstractions like qubits, gates, and circuits.
The compiler component translates high-level quantum code into low-level instructions optimized for execution on quantum hardware, often using intermediate representations like QASM~\cite{bishop2017qasm}. 
This process standardizes translation, applies optimizations, and reduces error-prone operations, which is crucial given the sensitivity of quantum systems to noise.
Finally, the execution environment supports running quantum programs on either actual quantum devices or simulators that emulate quantum conditions. 
Simulators are especially useful during development, as they allow testing under controlled conditions with noise models that approximate real hardware behavior~\cite{fingerhuth2018open}. 

\vspace{-4pt}
\subsection{Related Work}

Mining software repositories research has produced a vast array of studies across different domains, including machine learning, programming language adoption, mobile apps, and blockchain, among others. 
Research on language trends, such as~\cite{allamanis2013mining, ray2014large}, reveal how shifts impact software quality. 
Studies on bug prediction leverage historical data to improve defect prediction~\cite{kim2007predicting, rahman2013sample}. 
Eyolfson et al.~\cite{eyolfson2011time} explored how developer experience and commit timing affect bug introduction, shedding light on factors impacting software reliability. 
Mining repositories in Android development has exposed issues like API misuse and permission risks~\cite{li2018cid, alshehri2019puredroid}, while blockchain security research identifies common vulnerabilities in smart contracts~\cite{qian2022smart}. 
Additionally, Buse and Weimer~\cite{buse2010automatically} automated documentation from commit messages to enhance traceability. These studies exemplify how mining repositories capture critical insights from version control data to advance software quality, security, and development practices.

Existing studies on quantum software development primarily focus on small datasets, often limited to specific platforms or a narrow subset of quantum issues. For example, Paltenghi and Pradel~\cite{paltenghi-quantum-study} conducted an empirical study of bugs in quantum computing platforms by analyzing 223 bugs in 18 open-source projects, including major platforms such as Qiskit, Cirq, and Q\#. 
Although this study provides critical insights into quantum-specific bugs and recurring bug patterns, its scope remains limited to specific platforms, offering a narrower perspective on the quantum software landscape. 
Similarly, Zhao et al.~\cite{zhao2023identifying} collected a dataset of 36 bugs exclusively from the Qiskit platform, highlighting issues unique to quantum programming but restricting the analysis to a single framework.

This study seeks to fill this gap by analyzing a large dataset of GitHub repositories related to quantum computing. Through this analysis, we aim to provide a snapshot of the current state of quantum computing development, identify key trends, and offer insights into the collaborative dynamics within the community. 
Our findings will contribute to a deeper understanding of the landscape of quantum computing research and development and inform future efforts to advance the field.
\section{Methodology}
\label{sec:method}

\begin{table}[t]
    \centering
    \footnotesize
    \caption{File patterns used to extract package dependencies by programming language.}
    \vspace{-5pt}
    \begin{tabular}{|p{5em}|p{23em}|}
        \hline
        \textbf{Language} & \textbf{File Patterns} \\
\hline\hline
Python & \texttt{Pipfile}, \texttt{pyproject.toml}, \texttt{*requirements*.txt}, \texttt{requires*.txt},  \texttt{required*.txt}, \texttt{setup.py}, \texttt{setup.cfg} \\\hline
C++ & \texttt{CMakeLists.txt}, \texttt{Makefile}, \texttt{conanfile.txt} \\\hline
C & \texttt{CMakeLists.txt}, \texttt{Makefile} \\\hline
Lisp & \texttt{Makefile} \\\hline
Julia & \texttt{Project.toml} \\\hline
Rust & \texttt{Cargo.toml} \\\hline
    \end{tabular}
    \label{tab:language_file_patterns}
    \vspace{-10pt}
\end{table}

\subsection{Study Objective and Research Questions}
Following the Goal Question Metric (GQM) guidelines \cite{caldiera1994goal}, our research goal is structured as follows: Analyze \emph{quantum computing projects} for the purpose of \emph{exploring and categorizing} with respect to \emph{their current status, evolution, and maintenance} from the point of view of quantum computing researchers and developers in the context of GitHub repositories.

Two main research questions (RQ) arise from this goal. We explore quantum computing repositories to understand their development, popularity, and maintenance.

\begin{myquote}
\textbf{RQ1}. \textit{What is the current status and evolution of the Quantum Computing Community?}
\end{myquote}

\begin{itemize}
    \item RQ1.1: How has the popularity of quantum computing changed?
    \item RQ1.2: How have programming languages and framework usage trends evolved?
    \item RQ1.3: What is the developers' distribution for different frameworks and how do developers collaborate?
    \item RQ1.4: What trends and insights can be derived from topics and project categories?
\end{itemize}

\begin{myquote}
\textbf{RQ2}. \textit{How can we assess and classify the maintenance status of Quantum Computing repositories on GitHub based on their commit and issues data?}
\end{myquote}

\begin{itemize}
    \item RQ2.1: What commit patterns reveal over time?
    \item RQ2.2 How do the size and frequency of commits evolve over time?
    \item RQ2.3 What are the prominent topics in the reported issues?
    \item RQ2.4: How do various commit types (perfective, corrective, adaptive) influence the maintenance of repositories?
    \item RQ2.5: How can commit data be used to categorize the activity status of individual repositories?
    \item RQ2.6: How do repository characteristics vary across different activity levels?
\end{itemize}

\subsection{Dataset Construction}

\subsubsection{Data Collection}

We collected data from the Quantum Open Software Foundation (QOSF) list \cite{qosf-list} and expanded it using GitHub's Code Search API \cite{github-code-search-api}, focusing on package dependencies in standard Python project files (e.g., \texttt{requirements.txt}, \texttt{setup.py}) and keyword searches in relevant files (see Table~\ref{tab:language_file_patterns}). 
Due to the API's 1,000-result limit, we refined our search scope and manually inspected selected repositories. We also included repository forks, resulting in 95,011 repositories (2,799 originals and 92,212 forks).

\subsubsection{Preprocessing}

Since GitHub Code Search matches any substring, we need to make sure identified packages are genuinely used and not false positives. 
During the process, we manually inspected 1,300 search results to verify quantum-related repositories. 
We identified two common false-positive patterns: (1) repositories that copied content instead of forking and (2) monorepos containing both quantum-specific and unrelated libraries, such as the Azure SDK for .NET~\cite{azure-sdk}. 
To address this, we compiled a list of such repositories for exclusion.

Next, we shallow-cloned the repositories and analyzed standard files for package occurrences. 
We used \texttt{tokei} to detect programming languages and then searched for relevant package names in corresponding files (Table~\ref{tab:language_file_patterns}). 
We also maintained a mapping of programming languages to quantum framework package names (Table~\ref{tab:language_framework_patterns}). 
Since \texttt{tokei} does not detect certain languages like Q\#, we used GitHub Linguist~\cite{gh-linguist} as a fallback. This process yielded 24,374 repositories.

For commit analysis, we removed redundant commits from forks, focusing only on new developments. We also excluded tagged version commits (\texttt{ref: tag}), bot-authored commits, and anomalously high-volume commits. After multiple iterations of data cleaning and manual inspection, we retained 24,122 repositories and 1,232,828 commits from 10,697 users.

\begin{table}[t]
\footnotesize
\centering
\caption{Languages and their package names patterns.}
\vspace{-5pt}
\begin{tabular}{|p{4em}|p{22em}|}
\hline
\textbf{Language} & \textbf{File Patterns} \\
\hline\hline
Python & \texttt{cirq}, \texttt{amazon-braket-sdk}, \texttt{qiskit} \\\hline
C++ & \texttt{cuda-quantum}, \texttt{qpp} ,\texttt{qrack} \\\hline
C & \texttt{forest-sdk}, \texttt{qrack} \\\hline
Lisp & \texttt{forest-sdk}, \texttt{qcl} \\\hline
Julia & \texttt{PastaQ}, \texttt{Yao} \\\hline
Rust & \texttt{quil-rs} \\
\hline
\end{tabular}
\label{tab:language_framework_patterns}
\vspace{-10pt}
\end{table}

\subsection{Dataset Analysis}
Next, we describe the dataset and explain how various data points help address the research questions (RQs) outlined earlier. 
Our dataset includes 24,122 quantum computing repositories from GitHub, containing over 1.2 million commits, and 157,471 issues. 
These repositories cover diverse topics, such as simulation, quantum machine learning, algorithms, and frameworks, providing a foundation for analyzing the current state and evolution of quantum software.

\subsubsection{Repository Characteristics}
The dataset captures key information on project scale, community engagement, and activity levels, supporting RQ1 on the status and growth of quantum software. Typical repositories in the dataset have around 3.2 stars and 1.1 forks, though popular repositories like \texttt{Qiskit/qiskit} attract significantly more attention, with 5,167 stars and 2,353 forks, reflecting high community interest. 
Repositories contain 132,707 lines of code on average, contributing to a total of over 3.2 billion lines, underscoring the substantial codebase supporting quantum software.

\subsubsection{Issues and User Engagement}

The issue data provides insights for RQ2.3 which focuses on the issue patterns of the repositories.
The dataset includes 157,471 reported issues with 278,089 comments from 9,132 unique users, showing active community engagement in identifying and addressing challenges. 
Of these issues, 15,405 remain open, highlighting ongoing development needs. The issues cover both general software engineering concerns and quantum-specific topics, such as quantum experiments and matrix operations, helping us understand the unique demands within quantum software.

\subsubsection{Commit and Maintenance Analysis}
The commit data provides comprehensive details essential for answering RQ2.1, RQ2.2, RQ2.4, RQ2.5 and RQ2.6 on maintenance practices. With 1,232,828 commits, each entry includes information on the author, lines added or deleted, files changed, and commit timestamp, offering detailed insights into project evolution.
Using a fine-tuned DistilBERT model, we categorized commits into perfective, adaptive, and corrective types, allowing us to analyze how different maintenance actions influence software development. 
This classification reveals maintenance priorities, such as functionality enhancement, adaptability to new requirements, and bug resolution practices.

\section{Results}
\label{sec:results}

\subsection{Current Status and Evolution of Quantum Computing Community (RQ1)}

\subsubsection{How has Quantum Computing popularity changed?}
To evaluate the growth in popularity of quantum computing, we analyzed trends in both the number of repositories and the number of contributors over time.

\begin{figure}[t]
    \centering
    \vspace{-10pt}
   \includegraphics[width=0.96\linewidth]{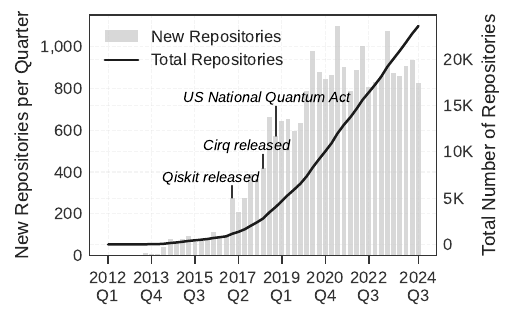}
   \vspace{-5pt}
    \caption{Quantum Computing repository growth over time.}
    \label{fig:repo_gowth}
    \vspace{-10pt}
\end{figure}

To find the number of quantum projects started, we analyzed the start dates of projects; this allows us to track the emergence of new repositories as well as the cumulative total of quantum computing projects. 
Figure~\ref{fig:repo_gowth} presents both the count of newly created repositories each quarter and the overall growth in the number of projects over time.
The data reveals a gradual increase in new projects from 2012 through 2016, reflecting early and moderate interest in quantum computing. 
However, around 2017, there is a noticeable acceleration in the number of new projects, suggesting a growing interest in quantum computing during this period. 
This increase becomes more prominent between 2018 and 2021, with especially sharp rises around 2019 and 2020. 
This surge is likely correlated with the release and adoption of accessible open-source quantum platforms, such as Qiskit (released in March 2017) and Cirq (released in July 2018), which lowered the barriers for new developers to enter the field.

By 2023, the cumulative number of quantum computing repositories nears 20,000, underscoring the field's expanding popularity and developer engagement. 
The steady increase in both new projects and total repository count reflects a rapidly growing ecosystem, indicating not only rising interest but also a commitment to developing resources and tools for quantum computing. 
This growth provides a strong indicator of quantum computing's increasing prominence in both academics and industry.

To further understand, how many developers are contributing to quantum projects, we analyze all the contributors to existing quantum projects and community growth over time. 

\begin{figure}[t]
    \centering
    \vspace{-10pt}
   \includegraphics[width=0.96\linewidth]{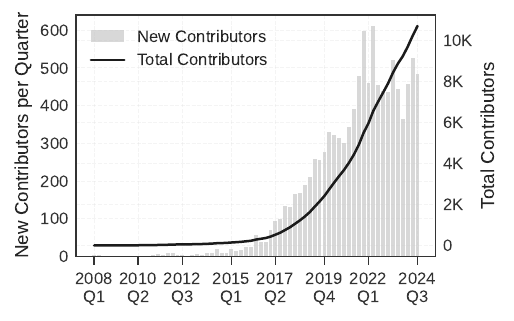} 
   \vspace{-5pt}
    \caption{Contributors' growth in Quantum computing projects.}
    \label{fig:repo_gowth}
    \vspace{-10pt}
\end{figure}

Complementing the repositories trend, the contributor growth (in Figure~\ref{fig:repo_gowth}) reveals a similar pattern, with new contributors joining steadily over time and a marked increase starting around 2018. 
The total number of contributors continues to grow, reaching 10,697 by 2024. 
This surge in active developers indicates a robust, expanding community contributing to quantum computing projects on GitHub.

\begin{myquote}
\textbf{Finding 1.1}. \textit{Quantum computing's popularity has shown a significant upward trend, as reflected in the steady increase in new repositories, total projects, and active contributors over time. The sharp rise beginning around 2017 underscores growing community engagement and sustained interest in quantum computing.}
\end{myquote}

\subsubsection{How have programming languages and framework usage trends evolved?}
To explore how programming language and framework usage trends have evolved in quantum computing, we examined project repositories over time, specifically focusing on the growth in the use of various languages and frameworks.

\begin{figure}[t]
    \centering
    \vspace{-10pt}
   \includegraphics[width=0.96\linewidth]{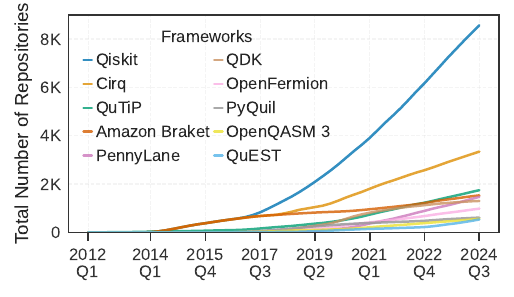} 
    \vspace{-10pt}
    \caption{Quantum computing framework usage over time, Qiskit remains the most popular framework.}
    \vspace{-15pt}
    \label{fig:framework_trends}
    
\end{figure}

\stitle{Framework Usage.}
Figure~\ref{fig:framework_trends} shows the evolution of popular quantum computing frameworks, with a strong dominance by Qiskit.
Starting around 2017, Qiskit~\cite{qiskitprogramming2021} rapidly gained traction, reflecting its widespread adoption and IBM's continued support and community engagement. 
Cirq~\cite{cirq} and QuTiP~\cite{qutip} also display steady growth, although at a slower pace compared to Qiskit. 
Notably, other frameworks, such as Amazon Braket SDK~\cite{amazon-braket} and PennyLane~\cite{pennylane}, began appearing around 2020, indicating newer entries into the quantum ecosystem, likely supported by commercial cloud providers like Amazon. 

This trend reflects a shift toward using comprehensive frameworks that simplify quantum computing development, offering specialized libraries and tools tailored to different quantum hardware and software needs. The rise in the variety of frameworks also underscores the diversification of quantum software development, with developers opting for tools that meet specific project or hardware requirements.

\begin{figure}[t]
    \centering
    \vspace{-10pt}
   \includegraphics[width=0.96\linewidth]{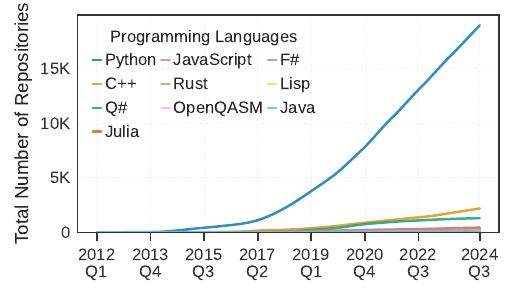}
    \vspace{-10pt}
    \caption{Usage of programming languages in Quantum repositories, Python remains the most popular language.}
    \vspace{-10pt}
    \label{fig:pl_trends}
\end{figure}

\stitle{Programming Languages Usage.}
Figure~\ref{fig:pl_trends} presents a clear trend toward Python as the most widely used language for quantum computing, consistent with the popularity of quantum frameworks such as Qiskit, Cirq, and PennyLane that support Python APIs. 
The widespread use of Python is evident, with a significant lead over other languages, illustrating its role as the primary language for quantum software development. 
C++ has a strong but comparatively smaller presence, probably because of its use in performance-critical components or low-level operations.

Emerging languages, such as Q\# from Microsoft~\cite{q-sharp} and OpenQASM~\cite{bishop2017qasm}, show steady growth around 2018, suggesting an increasing focus on languages explicitly designed for quantum applications. 
Julia and Rust have also begun to appear in recent years, reflecting a gradual diversification in language preferences as developers explore options that may offer performance benefits or specific features relevant to quantum computing.

\begin{myquote}
\textbf{Finding 1.2}. \textit{Programming language and framework usage in quantum computing has evolved notably, with Python and Qiskit dominating the landscape. However, recent growth in frameworks such as QuTiP and Amazon Braket SDK, along with emerging quantum-specific languages such as Q\# and OpenQASM, reflects a diversifying ecosystem tailored to specialized quantum applications.}
\end{myquote}

\subsubsection{What is the developers' distribution for different frameworks and how do developers collaborate?}

The distribution of developers across quantum computing frameworks reveals varying levels of reliance and engagement.  
As shown in Table~\ref{tab:dev_distribution}, Qiskit is the most widely adopted, with 8,561 repositories and 2,231 contributors relying on it for their projects, making it a central resource in the quantum computing ecosystem. 
Cirq and QuTiP follow, with 3,338 and 1,743 repositories, respectively, each supported by active communities of over 1,200 and 670 contributors. 
Other frameworks, such as PennyLane and Amazon Braket, are also widely utilized, with 1,464 and 1,525 repositories, backed by substantial contributor bases. 
Frameworks like OpenFermion and PyQuil, while linked to fewer repositories, maintain dedicated followings, suggesting that developers often choose frameworks based on specific applications or functionalities they offer.

\begin{table}[H]
\footnotesize
\centering
\caption{Distribution of repositories using different frameworks and associated contributors for repositories.}
\begin{tabular}{|l|r|r|}
\hline
\textbf{Framework} & \textbf{\# of repositories} & \textbf{\# of contributors} \\
\hline\hline
Amazon Braket & 1,525 & 292 \\\hline
Cirq & 3,338 & 1,205 \\\hline
OpenFermion & 979 & 318 \\\hline
OpenQASM 3 & 561 & 227 \\\hline
PennyLane & 1,464 & 558 \\\hline
PyQuil & 610 & 239 \\\hline
QDK & 1,299 & 286 \\\hline
Qiskit & 8,561 & 2,231 \\\hline
QuEST & 530 & 243 \\\hline
QuTiP & 1,743 & 673\\
\hline
\end{tabular}
\label{tab:dev_distribution}
\end{table}

\begin{myquote}
\textbf{Finding 1.3.1}. \textit{Qiskit is the most widely used quantum computing framework, followed by Cirq and QuTiP, with each framework attracting distinct developer communities based on specific functionalities and applications.}
\end{myquote}

\begin{figure}[t]
    \centering
    \includegraphics[width=0.96\linewidth]{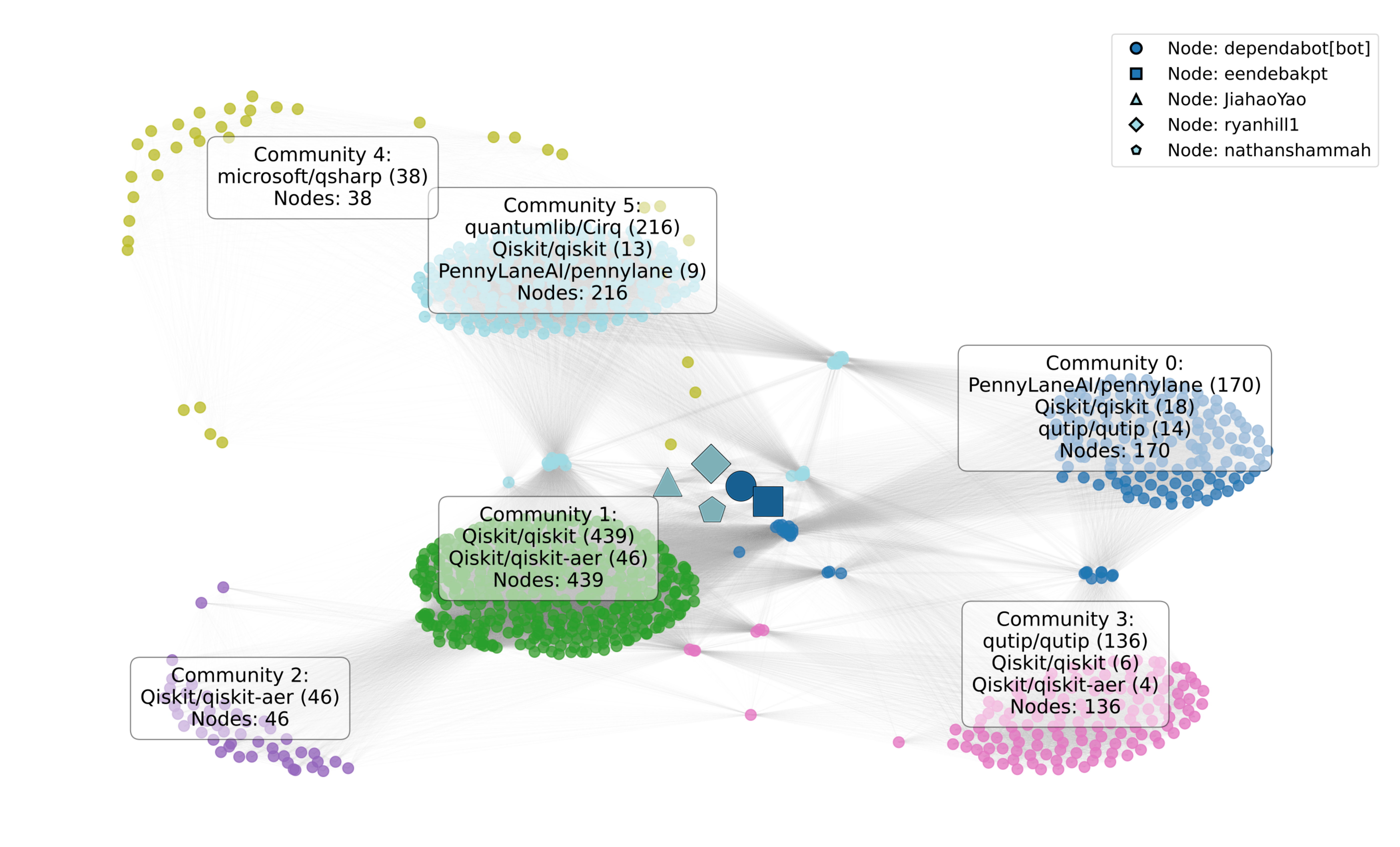} 
    \vspace{-10pt}
    \caption{Network of developers working with popular frameworks, highlighting the top 3 frameworks used in each cluster.}
    \label{fig:dev_network}
    \vspace{-15pt}
\end{figure}

To explore the collaboration among authors in popular quantum computing frameworks, we employ the Louvain algorithm~\cite{que2015scalable} to build a network of developers that have contributed to the most prominent frameworks found in RQ~1.2. 
We identify highly collaborative authors as those with significant participation across multiple frameworks.
Figure~\ref{fig:dev_network} illustrates the developers' network using six clusters, where each represents one of the prominent frameworks and highlights the top three frameworks within each cluster.
For brevity, we have only displayed the top-5 most collaborative authors, identified by the highest cross-framework participation, represented as the larger nodes in Figure~\ref{fig:dev_network}. 
The top-5 most collaborative authors, along with their respective cross-framework participation, are as follows: dependabot[bot] (5), eendebakpt (4), JiahaoYao (4), ryanhill1 (3), and nathanshammah (3). The most collaborative author in the network is a GitHub bot, dependabot[bot], due to its role in automating the management of project dependencies and ensuring they are consistently updated.

Figure~\ref{fig:dev_network} also shows the cross-framework collaboration in quantum computing. 
Specifically, the percentage of contributors who engage in cross-framework collaboration is 0\% for \texttt{Community 2}, 10\% for \texttt{Community 1}, 19\% for \texttt{Community 0}, 0\% for \texttt{Community 4}, 7\% for \texttt{Community 3}, and 10\% for \texttt{Community 5}.

\begin{table}[H]
\centering
\caption{Breakdown of developer backgrounds contributing  $\geq$3 prominent Quantum Frameworks.}
\vspace{-5pt}
\resizebox{\linewidth}{!}{%
\begin{tabular}{|c|c|c|c|c|c|}
\hline
\textbf{Researchers} & \textbf{Quantum Developers} & \textbf{Physicists} & \textbf{Grad. Students} & \textbf{Bot}  \\
\hline\hline
6.25\%  & 31.25\% & 18.75\% & 37.5\% & 6.25\% \\
\hline
\end{tabular}
}
\vspace{-10pt}
\label{tab:rq1.3-dev-background}
\end{table}

To further understand developers appearing as collaborators, we manually analyzed the GitHub profiles of the developers that contribute to three or more prominent frameworks. 
As shown in Table \ref{tab:rq1.3-dev-background}, the majority of contributors are graduate students (37.5\%), followed by quantum-focused developers (31.25\%), physicists (18.75\%), and researchers (6.25\%), with bots accounting for 6.25\%. This mix highlights a diverse community, where graduate students and specialized quantum developers are the most active collaborators, contributing essential expertise from both academia and applied quantum software development.

\begin{myquote}
\textbf{Finding 1.3.2}. \textit{
Collaboration patterns show limited cross-framework interaction, concentrated around a few key contributors. Graduate students and specialized quantum developers are the primary drivers of cross-repository collaboration.
}
\end{myquote}

\subsubsection{What trends and insights can be derived from project topics?}
\begin{figure}[t]
    \centering
   \includegraphics[width=0.8\linewidth]{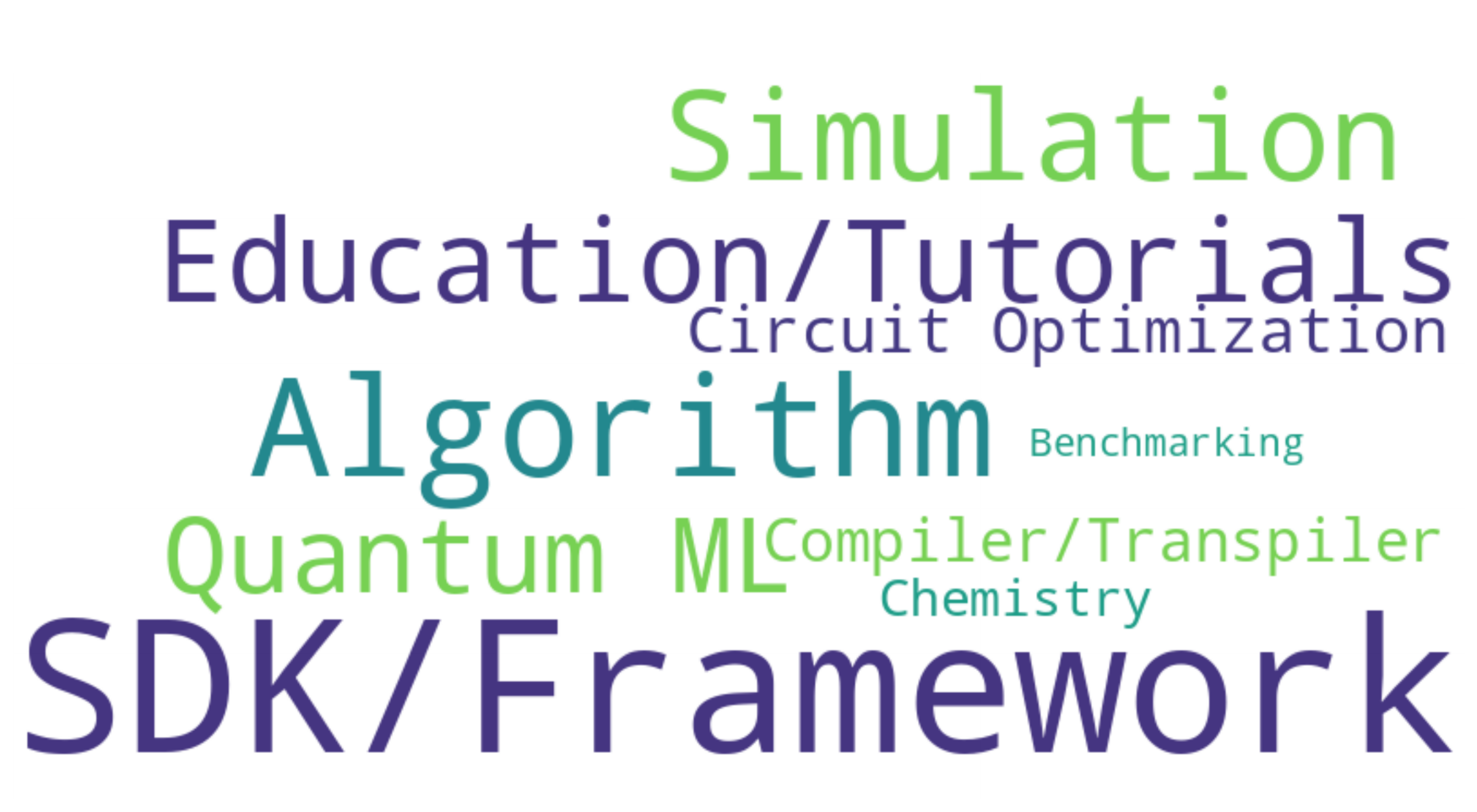} 
    \caption{Word cloud of topics from Quantum repositories.}
    \label{fig:repo_topics}
    \vspace{-15pt}
\end{figure}

\begin{figure}[H]
    \centering
   \includegraphics[width=0.96\linewidth]{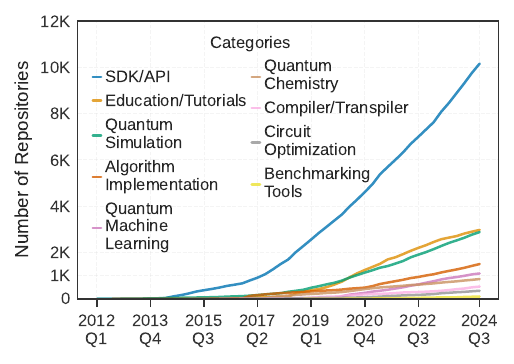} 
    \caption{Growth of different topics in Quantum repositories.}
    \label{fig:repo_topics_growth}
    \vspace{-10pt}
\end{figure}

To investigate project topics, we labeled repositories based on their primary focus (as shown in Figure~\ref{fig:repo_topics}) and analyzed trends in these topics over time. 
The timeline of repository growth across categories (in Figure~\ref{fig:repo_topics_growth}) reveals distinct patterns: SDK/API and Quantum Simulation projects exhibit the most substantial growth, indicating that toolkits and simulation capabilities are key areas of interest in the quantum computing field. 
Quantum Machine Learning and Algorithm Implementation also show steady growth, reflecting rising demand for quantum algorithms and their integration with machine learning.

The word cloud of repository categories further emphasizes the prominence of these areas, with terms related to Simulation, Machine Learning, and Compiler/Transpiler frequently appearing. 
This trend suggests a strong focus on tools that support both foundational quantum research and applied quantum computing projects, particularly in algorithm development and machine learning integration.

\begin{myquote}
\textbf{Finding 1.4}. \textit{The quantum computing community is not only focused on foundational frameworks and simulators but is also increasingly exploring applications in machine learning, chemistry, and advanced algorithm design.}
\end{myquote}

\subsection{Maintenance Analysis (RQ2)}
\subsubsection{What commit patterns reveal over time?}
\begin{figure}[t]
    \centering
    \vspace{-10pt}
   \includegraphics[width=0.96\linewidth]{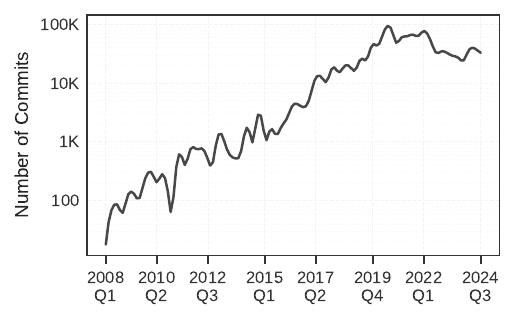} 
   \vspace{-5pt}
    \caption{Number of new commits in Quantum repositories show a consistent growth.}
    \label{fig:commit_pattern}
    \vspace{-10pt}
\end{figure}

We analyzed the overall commit activity in quantum computing repositories, as shown in Figure~\ref{fig:commit_pattern}. 
The graph illustrates commit counts overtime on a log scale, capturing trends from as early as 2008 through to 2024.

The data reveals a gradual increase in commits during the initial years, reflecting modest activity and slow growth in the quantum computing development landscape. 
However, starting around 2017, there has been a rise in commit frequency, likely driven by increased community engagement and the emergence of more accessible quantum development frameworks and tools. 
Again, similar to the repository growth, commit growth also significantly increases between 2020 and 2023, indicating increased development efforts, community contributions, and perhaps an expansion in the scope of quantum projects.

\begin{myquote}
\textbf{Finding 2.1}. \textit{The sharp rise in recent commit activity indicates that quantum computing is evolving from niche research into a collaborative, mature field, marked by consistent improvements and a growing number of contributors and projects.}
\end{myquote}

\subsubsection{How does the size and frequency of commits evolve?}
\begin{figure}[ht]
    \centering
   \includegraphics[width=0.96\linewidth]{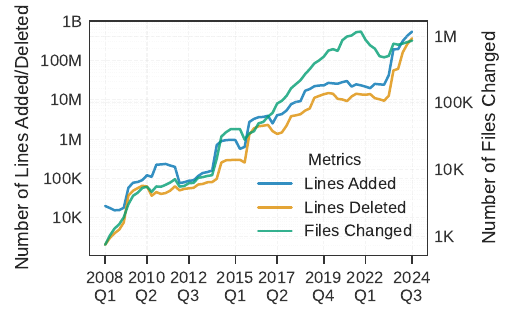} 
   \vspace{-10pt}
    \caption{Code changes trend in Quantum software, in terms of the number of lines added, deleted, and files changed.}
    \label{fig:changes_trend}
    
\end{figure}

\begin{figure}[t]
    \centering
    \vspace{-10pt}
    \includegraphics[width=0.96\linewidth]{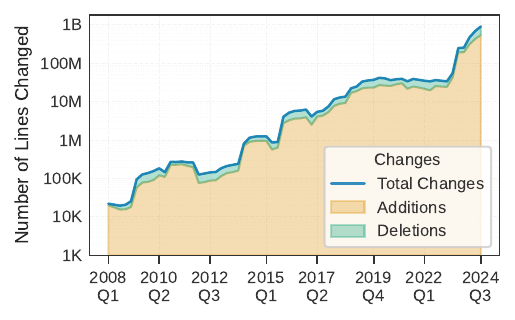} 
   \vspace{-10pt}
    \caption{Volume of code changes in Quantum repositories, the number of lines added, deleted, and files changed.}
    \label{fig:code_volume}
    \vspace{-10pt}
\end{figure}

We examine the trends in code volume and file changes. Figure~\ref{fig:changes_trend} and Figure~\ref{fig:code_volume} show a steady increase in both the number of lines added/deleted and files modified, with a significant spike around 2020-2023.

This surge in file changes corresponds to a noticeable jump in activity due to large-scale code reformatting, which temporarily elevated the number of files modified. 
This pattern aligns with the previously observed trend of accelerated commit activity post-2017, reflecting a period of heightened development and restructuring in quantum computing projects. 
These spikes indicate both growing project complexity and increased focus on code organization and standardization as the field matures.

\begin{myquote}
\textbf{Finding 2.2}. \textit{Commit size and frequency have risen significantly, with a spike from large-scale code reformatting around 2020-2023, indicating increasing project complexity and a move toward standardized development in quantum computing.}
\end{myquote}

\subsubsection{What are the prominent topics in the reported
issues?}

\begin{table*}[t]
\footnotesize
\centering
\caption{Description of the key features in the issues for Quantum Computing repositories.}
\vspace{-5pt}
\begin{tabular}{|l|l|c|}
\hline 
\textbf{Topic} & \multicolumn{1}{c}{\textbf{Components}} & \textbf{Frequency(\%)} \\ \hline \hline
Dependency Management and Updates & deprecate, remove, bump, version, dependency, update, pypi, pip & 14.76 \\ \hline
Quantum Computing Terminology & qubit, quantum, pulse, hamiltonian, matrix, gate, pauli, phase, rotation & 11.33 \\ \hline
Code Maintenance and Refactoring & rename, update, refactor, fix, remove, typo, change & 9.85 \\ \hline
Quantum Computing Software and Frameworks & qasm, qaoa, openqasm, qobj, qnode, qpu, qiskit, qutip, ibmq & 7.94 \\ \hline
Fixing Bugs & error, fix, fail, bug, warning, log, issue & 6.44 \\ \hline
Documentation & doc, documentation, readme, docstring, tutorial, docs, sphinx & 5.75 \\ \hline
Quantum Computing Hardware Interfaces & driver, backend, device, channel, job, add, instrument, executor, gpu & 5.72 \\ \hline
Continuous Integration and Testing & test, ci, travis, coverage, build, workflow, run, testing, integration & 5.68 \\ \hline
Quantum Computing Experiments & measurement, calibration, benchmark, performance, experiment, noise & 5.59 \\ \hline
Version Control & merge, update, release, link, branch, master, version, main, pr & 4.66 \\ \hline
Matrix Operations & state, gradient, solver, matrix, vqe, observable, adjoint, sparse & 4.08 \\ \hline
Python Syntax & import, module, json, dataset, serialization, kwarg, decorator & 2.66 \\ \hline
Data Visualization and Machine Learning Libraries & matplotlib, tensor, graph, plot, tensorflow, notebook, lattice, torch, visualization & 2.16 \\ \hline
Other & \begin{tabular}[c]{@{}l@{}}circuit, pass, drawer, transpiler, register, operator, wire, datum, seed, simulator,\\ rb, shot, sampler, optimizer, numpy, python, rust\end{tabular} & 13.37 \\ \hline
\end{tabular}
\vspace{-15pt}
\label{tbl:issue_topics}
\end{table*}

We identified key topics in quantum computing repository issues using BERTopic~\cite{grootendorst2022bertopic}. However, because unsupervised BERTopic can generate noisy or inconsistent topics that are difficult to interpret, we adopted a semi-supervised approach to improve its reliability. Specifically, we initialized BERTopic with seed topics -- keywords selected based on their high TF-IDF scores -- to guide the model toward coherent and meaningful clusters rather than arbitrary groupings.

To assess the effectiveness of our approach, we conducted a qualitative validation focusing on two key aspects: (1)~whether BERTopic formed coherent and meaningful topics, and (2)~whether it generalized beyond the explicit seed terms. We manually reviewed a diverse subset of issues from each cluster to evaluate whether BERTopic not only grouped semantically similar issues in the same cluster -- even when they did not explicitly contain the seed topic keywords -- but also identified a broader range of topics beyond those initially defined by the seed terms. 
Our qualitative analysis confirmed that the seeded BERTopic approach facilitated the discovery of meaningful clusters beyond the original TF-IDF terms while improving topic cohesion and reducing noise.
This approach enabled us to uncover various issue categories relevant to quantum software development, summarized in Table~\ref{tbl:issue_topics}.

Our analysis revealed that a significant portion of issues (around 47.14\%) are related to classic software engineering challenges, including dependency management, code maintenance, refactoring, documentation, continuous integration and testing, and version control. 
These issues reflect common needs in software projects, such as managing package versions, fixing bugs, improving code structure, and ensuring seamless integration and deployment processes. 
The prevalence of these topics underscores that, like other software fields, quantum computing projects require robust tools and practices for maintaining code quality and collaboration.

About 34.66\% of the reported issues are specific to quantum computing, focusing on specialized topics such as quantum computing basics (e.g., qubits, rotation), hardware interfaces, quantum computing experiments, matrix operations, and quantum-specific software frameworks. 
These issues highlight the unique challenges in quantum software, including the need for tools that enable the simulation of quantum experiments and support the complex mathematical operations essential for quantum algorithms.

\begin{myquote}
\textbf{Finding 2.3}. \textit{Approximately 47\% of issues in quantum computing repositories relate to classic software engineering challenges, while 34\% are quantum-specific like quantum experiments and matrix operations. This indicates a need for specialized tools to address the unique requirements of quantum software development, as well as a demand for overall robust software infrastructure to support quantum computing.
}
\end{myquote}

\subsubsection{Types of Commits}
Understanding the proportions of different types of commits is important to understand advances in quantum computing.
To identify the different commit types in our repositories, we used the DistilBERT model fine-tuned by~\cite{hf-ml-msr} on GitHub commit messages~\cite{sarwar2020multi}. 
This model classified the commit messages into categories based on their purpose: perfective (improvements), adaptive (adjustments to changes), and corrective (bug fixes).

As shown in Table \ref{table:classification_commit}, the majority of commits are perfective (51.76\%), this finding reflects an emphasis on improving and refining code to enhance functionality and performance. 
Adaptive commits, at 22.58\%, indicate frequent adjustments to accommodate new requirements or technologies, while corrective commits, at 18.54\%, suggest a relatively lower occurrence of bug fixes. 
The presence of combined categories, such as adaptive-perfective and corrective-perfective, suggests overlapping maintenance efforts, where developers enhance functionality while adapting or fixing the code.
\begin{table}[H]
\centering
\vspace{-10pt}
\footnotesize
\caption{Percentage of Frequency for Each Classification}
\begin{tabular}{|l|r|}
\hline
\textbf{Classification} & \textbf{Percentage (\%)} \\
\hline\hline
Perfective                      & 51.76 \\
Adaptive                        & 22.58 \\
Corrective                      & 18.54 \\
Adaptive Perfective              & 4.97  \\
Corrective Perfective            & 1.23  \\
Corrective Adaptive              & 0.46  \\
\hline
\end{tabular}
\label{table:classification_commit}
\vspace{-10pt}
\end{table}

\begin{myquote}
\textbf{Finding 2.4}. \textit{The majority of commits in quantum computing repositories focus on code improvements, followed by adjustments, while corrective commits are less frequent. This suggests an emphasis on enhancing functionality over bug fixing, with potential gaps in tracking and addressing bugs. }
\end{myquote}

\subsubsection{Classification of Repository Activity-Levels using Commit Data}
\begin{figure}[t]
\centering
\vspace{-10pt}
\includegraphics[width=0.96\linewidth]{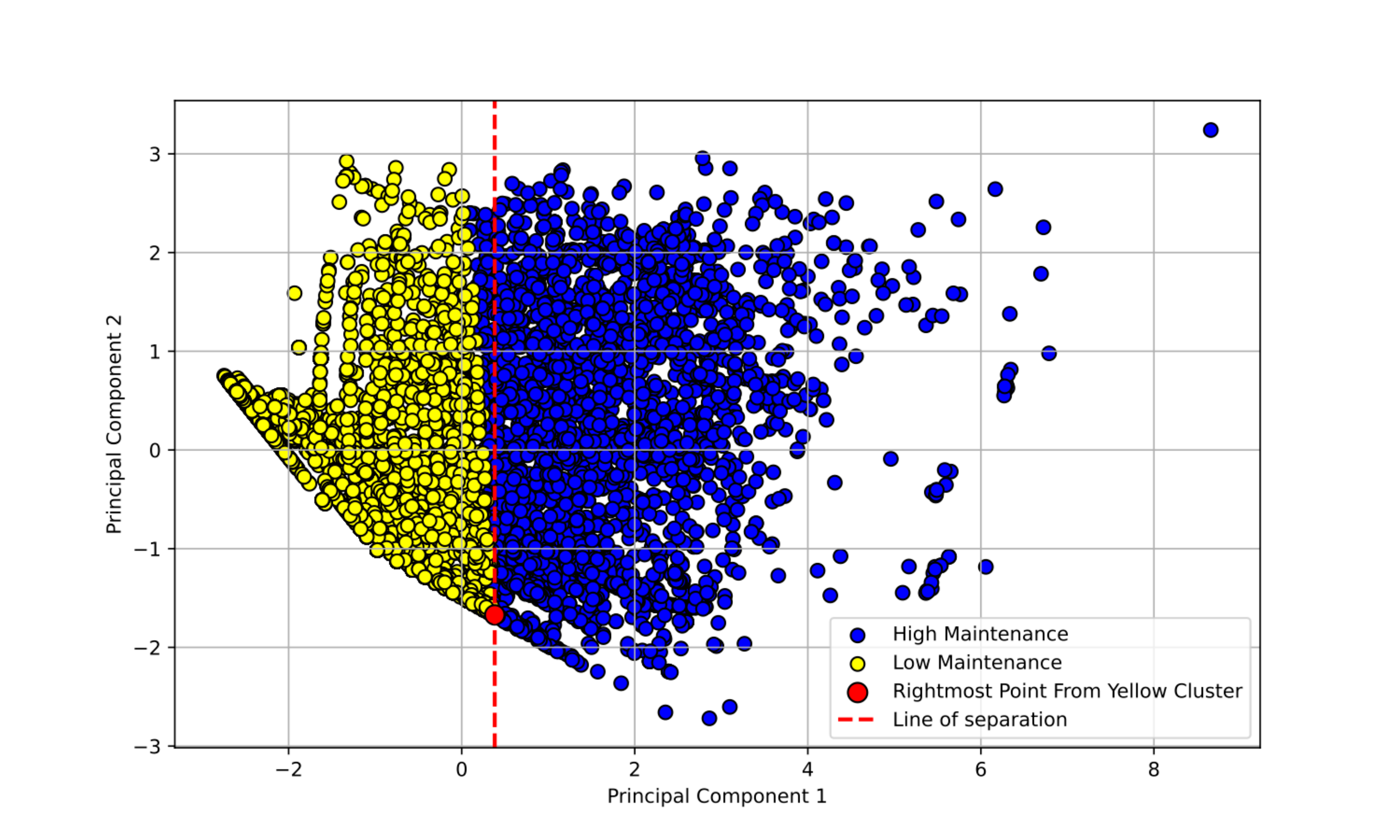} 
\vspace{-10pt}
\caption{Clustering Quantum repositories into 2 classes: high and low activity repositories.}
\label{fig:rq2.5-kmeans}
\vspace{-15pt}
\end{figure}

To account for potential duplication in forked repositories, we analyzed only original commits not present in parent repositories, resulting in 8,548 repositories being included in our analysis. We examine the factors that differentiate high-activity repositories from those with lower activity through K-means clustering~\cite{lloyd1982least,macqueen1967some} of repository commit attributes, including commit frequency per month, average gap between commits, number of commits, number of contributors, and project duration.
Specifically, we apply K-means clustering to assign labels to the repositories and then perform 2D Principal Component Analysis (PCA) ~\cite{hotelling1933analysis,jolliffe2005principal} to visualize the clusters, as shown in Figure~\ref{fig:rq2.5-kmeans}.
Figure~\ref{fig:rq2.5-kmeans} illustrates that the dashed red line, passing through the rightmost lower-activity point (red circle), effectively separates high-activity repositories (blue) from lower-activity ones (yellow).
Additionally, we observe that high-activity repositories tend to exhibit greater variance, as evidenced by more spread-out scatter points and outliers.

\begin{table}[H]
\centering
\vspace{-5pt}
\caption{Mean centroids of repository activity categories.}
\resizebox{\linewidth}{!}{%
\begin{tabular}{|l|c|c|c|c|c|}
\hline
\textbf{Category} & \textbf{Frequency} & \textbf{Avg. Days b/w Commits} & \textbf{Commits} & \textbf{Authors} & \textbf{Months} \\
\hline\hline
High  &20.05 &4.65 & 187.838 & 4.523 & 3.9 \\
\hline
Low  & 2.86 & 387.535 & 9.351 & 1.296 & 1.47 \\
\hline
\end{tabular}%
}
\label{table:rq2.5-kmeans}
\vspace{-10pt}
\end{table}

Table \ref{table:rq2.5-kmeans} displays the attributes of the centroids for the two clusters, highlighting that high-activity repositories are more actively managed, as evidenced by their higher commit frequency, shorter commit gaps, larger number of commits, more contributors, and longer duration of maintenance.

\begin{myquote}
\textbf{Finding 2.5}. \textit{Highly active quantum repositories have more contributors and are sustained over longer periods. However, there are fewer high-activity repositories (3,562) compared to those with low-activity (4,986), highlighting the disparity in the level of development and support in the quantum computing ecosystem.}
\end{myquote}

\subsubsection{How do different repository characteristics differ between maintenance activity levels?}
To explore differences between high- and low-activity repositories, we examined various attributes and commit categories, using the Mann-Whitney U Test for statistical significance~\cite{nachar2008mann}. 

\begin{table}[H]
\centering
\caption{Mann-Whitney U Test results for continuous variables.}
\resizebox{\linewidth}{!}{
\begin{tabular}{|l|c|c|l|}
\hline
\textbf{Variable} & \begin{tabular}{@{}c@{}}\textbf{High Activity} \\ \textbf{Mean}\end{tabular}  & \begin{tabular}{@{}c@{}}\textbf{Low Activity} \\ \textbf{Mean}\end{tabular} & \textbf{p-value} \\
\hline\hline
Update Duration & 281.9597 & 147.6592 & \textless{}0.001 \\
Stars & 6.5573 & 0.61462 & \textless{}0.001 \\
Forks & 2.1212 & 0.20753 & \textless{}0.001 \\
Size & 40581.0595 & 32940.8206 & \textless{}0.001 \\
Contributors & 53.2758 & 62.6904 & 0.044 \\
Lines of code & 468798.7614 & 115685.0642 & \textless{}0.001 
\\ \hline
\end{tabular}
}
\label{tab:u-test}
\vspace{-5pt}
\end{table}

The results in Table~\ref{tab:u-test} reveal notable distinctions between these repository types. 
High-activity repositories show a significantly longer update duration, averaging 281.96 days compared to 147.66 days in low-activity repositories (p $<$ 0.001), which suggests more consistent, ongoing updates over time. 
They also tend to attract more stars and forks, with high-activity repositories averaging 6.56 stars and 2.12 forks versus 0.61 stars and 0.21 forks for low-activity ones, indicating greater community interest and engagement. 
In terms of size, high-activity repositories have substantially more lines of code (468,798 vs 115,685) and overall project size, reflecting more complex codebases that likely demand intensive upkeep. 
Interestingly, low-activity repositories have a slightly higher mean contributor count (62.69 vs 53.28, p = 0.044), suggesting low-activity projects may receive occasional contributions without sustained commitment.

\begin{figure}[t]
    \centering
    \vspace{-10pt}
   \includegraphics[width=0.96\linewidth]{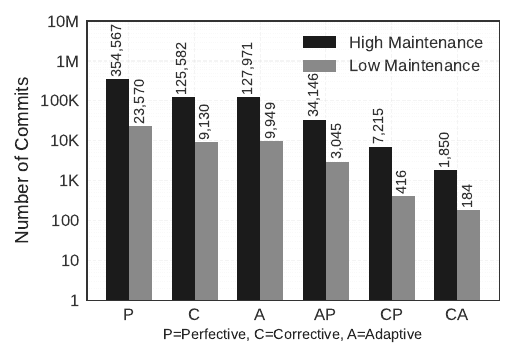} 
   \vspace{-10pt}
    \caption{Distribution of different commit classes for high and low activity repositories.}
    \vspace{-15pt}
    \label{fig:high_low_commits}
\end{figure}

The distribution of commit types across activity levels, shown in Figure~\ref{fig:high_low_commits}, further supports these findings. 
High-activity repositories show a higher frequency of perfective and adaptive commits, focusing on ongoing improvements and adaptations to changes, while low-activity repositories exhibit fewer updates across all commit categories. 
These insights highlight that high-activity repositories are generally larger, attract more engagement, and emphasize continuous enhancement and adaptability, while low-activity repositories may experience sporadic contributions without regular development efforts.

\begin{myquote}
\textbf{Finding 2.6}. \textit{High-activity repositories have frequent updates, larger codebases, and greater community engagement, while low-activity repositories show irregular updates and smaller codebases.}
\end{myquote}

\section{Implications}
\label{sec:implications}
Our findings highlight key challenges in quantum software engineering and outline actionable steps to improve development practices, tooling, and education.

\begin{enumerate}[label=\emph{(\alph*)}]
    \item \emph{Bug Resolution and Software Reliability.} The prevalence of perfective commits (51.76\%) over corrective commits (18.54\%) suggests that defect resolution is underemphasized. Prior work by Paltenghi and Pradel~\cite{paltenghi-quantum-study} found that quantum software exhibits recurring bug patterns, yet our findings indicate that corrective actions remain limited. Enhancing automated testing, issue-tracking systems, and debugging frameworks is critical for improving software reliability.

    \item \emph{Quantum-Specific Tooling Needs.} With 34\% of reported issues being unique to quantum computing, there is a clear need for quantum-aware debugging tools, enhanced simulators, and automated static analysis techniques. Zhao et al.~\cite{zhao2023identifying} identified debugging as a major challenge in Qiskit, and our findings reinforce the necessity for specialized tooling to address quantum-specific issues, such as circuit optimization and error mitigation.

    \item \emph{Sustaining Software Development.} The divide between \emph{high- and low-activity repositories} indicates challenges in long-term software maintenance. Studies on repository evolution~\cite{quantum-issues, zappin2024quantum} suggest that engagement in quantum projects is often short-lived, which aligns with our observation that many repositories exhibit sporadic contributions. Encouraging community-driven contributions, better documentation practices, and sustainable governance models can improve repository longevity.

    \item \emph{Standardization and Interoperability.} As quantum computing frameworks evolve, ensuring consistent APIs, cross-framework compatibility, and adherence to software engineering best practices will enhance usability and reduce fragmentation. Prior work on quantum programming languages~\cite{zhao_qml} has shown that the ecosystem remains highly fragmented, and our findings confirm that while Python and Qiskit dominate, newer frameworks, such as Cirq and PennyLane are gaining traction, emphasizing the need for standardization.

\end{enumerate}

To drive advancements in quantum software engineering, we open-source our dataset, providing the broader software engineering community with empirical insights to inform tool development, maintenance strategies, and best practices in quantum software development.

\section{Threats to Validity}
\label{threats}
\stitle{Internal Validity.}
A potential threat to internal validity lies in the criteria used for selecting quantum-related repositories. 
While we aimed to ensure relevance by filtering for repositories with quantum-specific keywords and technologies, the initial reliance on GitHub metadata alone may have introduced some irrelevant or incomplete projects. 
To mitigate this, we performed extensive data cleaning as described in the processing section, where we removed unchanged forks, retained only repositories demonstrating meaningful activity, and ensured the presence of quantum package usage. 
These steps helped refine the dataset to better represent genuine quantum software development projects. 
However, there remains a risk that some valid but low-activity repositories could have been excluded, possibly omitting some niche or experimental quantum projects.

\stitle{External Validity.}
Our findings may have limited generalizability beyond GitHub to other platforms where quantum software may be developed, such as proprietary or institution-hosted repositories. 
This study focuses exclusively on open-source projects on GitHub, which could lead to insights that reflect primarily the open-source quantum development community. 
This limitation suggests that proprietary or enterprise-level practices may not be fully captured here, as they might involve different workflows, tools, or challenges. 
Thus, while the results provide valuable insights into open-source quantum software, extending the study to include other sources could yield a broader understanding of quantum software practices.

\stitle{Construct Validity.}
The construct validity of this study could be impacted by limitations inherent to mining software repositories. 
Metrics such as commit counts, issue reports, and collaboration structures provide quantitative insights into development activities but may overlook qualitative factors such as developer expertise, code quality, or algorithm complexity. 
The absence of standardized best practices and tools in quantum programming complicates the interpretation of certain development patterns and issues. 
To address this, we incorporated additional filtering and categorization criteria based on quantum package usage, commit activity, and algorithm types, helping to focus the dataset on relevant aspects of quantum software engineering.

\stitle{Reliability.}
The reliability of our findings is based on the reproducibility of the data mining and cleaning techniques applied. 
Future changes to GitHub's API or to repository metadata could impact the replicability of aspects of our methodology. 
Additionally, reliance on automated processes like natural language processing (NLP) for issue and commit classification may introduce some errors in categorization. 
To reduce this risk, we detailed the data extraction and cleaning process and cross-validated classifications where feasible to ensure consistency.

By recognizing these limitations, we aim to provide transparency about the study's scope and the validity of our findings within this context.

\section{Conclusions}

This study provides a comprehensive analysis of the current state, evolution, and maintenance practices in the quantum computing community by examining over 21,000 GitHub repositories, containing more than 1.2 million commits from over 10,000 unique contributors.
Through this large-scale mining effort, we evaluate community growth, programming trends, and maintenance practices as well as offer insights into the strengths and areas for improvement in quantum software development.
Our findings show a rapidly growing community, with a 200\% increase in the number of repositories and a 150\% rise in contributors since 2017, underscoring the growing interest in quantum computing. However, our commit analysis reveals a focus on perfective updates, with fewer corrective commits, which suggests potential gaps in bug resolution. Additionally, a third of the issues emphasize the need for specialized tools alongside general software infrastructure.

Based on our analysis of development patterns, community structure, and maintenance practices, we recommend investment in specialized tooling and systematic maintenance frameworks for quantum computing. 
Enhanced support for high-maintenance repositories could drive further innovation. 
Additionally, extending this analysis to other platforms may provide broader insights into the quantum software landscape.

\balance
\bibliographystyle{plain}

\bibliography{IEEEexample.bib}

\end{document}